\documentclass[twocolumn,showpacs,amsmath,amssymb,prl]{revtex4}
\usepackage{graphicx}% Include figure files
\usepackage{soul}
\usepackage{color}

\begin{document}
\title{Ratcheting Heat Flux against a Thermal Bias}
\author{Nianbei Li$^1$}
\author{Peter H\"anggi$^{1,2}$}
\email{Hanggi@Physik.Uni-Augsburg.DE}
\author{Baowen Li$^{1,3}$}
\email{phylibw@nus.edu.sg} %\altaffiliation {Corresponding author}
\address{$^1$ Department of Physics and Centre for Computational Science
         and Engineering, National University of Singapore, Republic of Singapore
         117542\\
         $^2$ Institut f\"ur Physik, Universit\"at Augsburg Universit\"atsstr. 1
D-86135 Augsburg, Germany\\
         $^3$ NUS Graduate School for Integrative Sciences and Engineering,
         Singapore 117597, Republic of Singapore}

%\date{4 March, 2008}

\begin{abstract}
Merely  rocking  the temperature in one heat bath can direct a
steady heat flux   from cold to hot against a (time-averaged)
non-zero thermal bias in  stylized nonlinear lattice junctions that
are sandwiched between two heat baths. Likewise, for an average
zero-temperature difference between the two contacts a net,
ratchet-like heat flux emerges. Computer simulations show that this
very heat flux can be manipulated and even {\it reversed} by
suitably tailoring the frequency ($\lesssim$ 100 MHz) of the
alternating temperature field.
\end{abstract}

% 07.20.Pe Heat engines; heat pumps; heat pipes
% 44.90.+c Other topics in heat transfer
%85.90.+h Other topics in electronic and magnetic devices and microelectronics
%05.90.+m Other topics in statistical physics, thermodynamics, and
%nonlinear dynamical systems (restricted to new topics in section
%05)
%05.40.-a   Fluctuation phenomena, random processes, noise, and Brownian motion

\maketitle

The generation of heat flow and its controlled manipulation presents
an ever-growing endeavor for mankind. In quest of  its technological
solution we could witness substantial progress over the least
decades, with first serious efforts being achieved that can be
traced back to the early 1960's \cite{CStarr, Williams66, Eastman68,
Thomas70, rectifier70}. This underlying challenge does not present
plain sailing because phonons are by far more difficult to control
than electrons and photons. The recent years, however, have given
headway to new advances. In particular, thermal rectifiers have been
designed theoretically
\cite{rectifier,diode,quantumdiode,Hu06,peyrard,yang} with a first
experimental realization put forward with help of asymmetric
nanotubes \cite{experimentaldiode}. Moreover, using the
 negative differential thermal resistance \cite{diode}, a
thermal transistor  has been proposed \cite{transistor}, which is
able to control heat flow much like a Field-Effect-Transistor(FET)
does for electric currents. Even different thermal logic gates
\cite{WangLi07} have been conceived. All this progress implies that
phonons, -- traditionally being regarded rather as a nuisance --,
can in fact be put to work constructively in order to carry and
process information effectively. Altogether, this has giving cause
for a new discipline - phononics -, i.e.  the science and
engineering of phonons \cite{WangLi08}.

In addressing this theme let us remind again of the original
formulation of the second Law by Rudolf Clausius in 1850 which
states that heat cannot spontaneously flow from a subsystem at lower
temperature to a coupled subsystem at higher
temperature$^1$\footnotetext[1]{The correct formulation of the
second law involves quantities at (constrained) thermal equilibrium
only; in particular, no time variable $t$ enters the formulation of
the 2nd  Law \cite{Thompson}}. Thus, in order to generate a steady
heat flow against a thermal bias, or even generate heat flow in
absence of a thermal bias, the system {\it necessarily} must operate
{\it away} from thermal equilibrium, beyond the limiting realm of
the $2$nd Law. A typical such situation that comes to mind is the
Peltier effect where a steady heat flow is generated via imposing a
stationary electric current across an isothermal junction of two
different materials.

\begin{figure}
\includegraphics[width=\columnwidth]{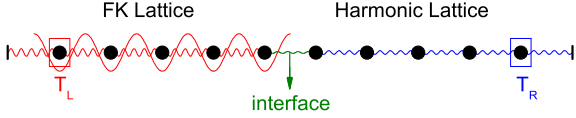}
\vspace{-.5cm} \caption{(Color-online)  Schematic setup of a weakly
coupled Frenkel-Kontorova lattice to a Harmonic lattice, being
coupled to two heat baths at temperatures $T_L(t):=T_L$ and $T_R$.
%(b) Heat
%flux $J$ {\it vs.} thermal bias $\Delta$ at vanishing  driving
%($A=0$)for different lattice lengths $N=50+50$ and $N=100+100$. The
%average temperature is $T_0=0.09$ for both cases.
} \label{fig:1}
\end{figure}

With this work we  propose  via computer simulations an intriguing
new scheme that does not require the resource of a stationary
non-equilibrium bias in the form of e.g. a stationary electric
current but instead combines the elements of an asymmetric lattice
structure with a {\it non-biased}, temporally alternating bath
temperature. Dwelling on ideas from the field of Brownian motors
\cite{BM1,BM2,BM3,BM4}, -- originally devised for Brownian particle
transport, -- we here attempt to direct a priori energy (heat)
across a spatially extended nonlinear lattice, see Fig. \ref{fig:1},
against an external thermal bias.
%%%%%%%% new %%%%%%%%%%%%%%
This objective is thus similar in spirit for devising machines and
devices that can pump heat on a molecular scale
\cite{vandenbroeckPRL2006,segalPRE2006,marathePRE2007,vandenbroeckPRL2008}.
%%%%%%%%%%%
In
doing so, the lattice system is brought into contact with two heat
baths, with one bath subjected to a time-varying temperature.
%The latter thermal contact then allows for
%a control of heat flow across this so designed open system.
Taken alone, this nonlinear lattice structure exhibits a thermal
diode effect \cite{rectifier,diode} which can be exploited to
function as a heat ratchet device when an additional source of
nonequilibrium, -- here realized with a rocking bath temperature --
, is present. Then, merely rocking the  temperature in one heat bath
can induce {\it dynamically} a finite bias between the two heat
baths, being held at the same time-averaged temperature. This novel
nonequilibrium ratchet feature can be utilized (i) to reverse the
flux, to (ii) direct heat flow from cold to hot against an average
thermal bias, or even to (iii) turn a regime with a {\it negative}
differential thermal resistance (NDTR) into a regime with positive
DTR, and vice versa.

%Figure 1

Explicitly, we study numerically a system composed of a
Frenkel-Kontorova (FK) nonlinear lattice of period $a$ which is
weakly coupled to a Harmonic lattice (HL), each consisting of $N/2$
atoms of identical mass $m$. This  setup is shown in Fig.
\ref{fig:1}, with the FK lattice on the left and Harmonic lattice on
the right. The FK-HL configuration is governed by the Hamiltonian:
\begin{eqnarray}
H=\sum^{\frac{N}{2}}_{i=1}\left[\frac{p^2_i}{2m}+
\frac{k_L}{2}(q_i-q_{i-1})^2-\frac{V_L}{(2\pi)^2}\cos \frac{2\pi q_i}{a}\right]\nonumber\\
+\frac{k_{int}}{2}(q_{\frac{N}{2}+1}-q_{\frac{N}{2}})^2+
\sum^{N}_{i=\frac{N}{2}+1}\left[\frac{p^2_i}{2m}+\frac{k_R}{2}(q_{i+1}-q_{i})^2\right]
\end{eqnarray}

Herein, $q_i=x_i -ia$ denotes the displacement from equilibrium
position $ia$ for $i$-th atom, $a$ is the lattice period, $k_L$ and
$V_L$ are the spring constant and the on-site potential of the FK
lattice, $k_{int}$ is the coupling strength between the FK and the
Harmonic lattice, and $k_R$ is the spring constant of the Harmonic
lattice. Fixed boundary conditions, yielding $q_0=q_{N+1}=0$, have
been employed. The $1$-st atom  and the $N$-th atom  are put into
contact with two Langevin heat baths possessing temperature
$T_L(t):=T_L$ and $T_R$, respectively. Gaussian white noise are
used, namely, $\langle\xi_{1/N}(t)\rangle=0$ and
$\langle\xi_{1/N}(t)\xi_{1/N}(0)\rangle=2k_B\eta T_{L/R}\delta(t)$.
$k_B$ is the Boltzmann constant and $\eta$ denotes the coupling
strength between system and heat bath. The time-varying  heat bath
temperature $T_{L}(t)$, oscillates dichotomously at angular
frequency $\omega$ and driving strength $A$. The used bath
temperatures thus read explicitly:
\begin{eqnarray}\label{T-setup}
T_L(t) := T_{L}&=&T_0(1+\Delta+A\cdot \text{sgn}(\sin{\omega t})),\nonumber\\
T_R&=&T_0(1-\Delta)\;,
\end{eqnarray}

where $T_0=(\overline{T_L(t)}+T_R)/2$ is the temporally averaged
environmental reference temperature, $2\Delta=(\overline{T_L(t)}-T_R)/T_0$ denotes the
normalized temperature difference, and $\text{sgn}(\sin{\omega t})$
provides the dichotomous, time-dependent temperature variation.
%%%%%%%%%%%%% NEW
The time scale  $\omega$ of the temperature manipulation of the heat
bath is assumed to vary much slower than the time scale $\tau_{leq}$
to reach local thermal equilibrium; i.e. $\omega <<
\tau^{-1}_{leq}$. This time scale  for good thermal conductors is
typically a  function of temperature; it is of the order of the time
scale of the electron-phonon relaxation time which normally
decreases  with decreasing temperature. For good metals this time
scale is of the order of $0.1-1$  picoseconds.
%%%%%%%%%%%%%%%%%%%%%%%%%%%%%%%%%%%%%%%%%%%%%%%%%%%%%%%%%%%%%%%%%%%%%%%%%%%%%%%%%%%%%

We next use dimensionless parameters by measuring positions in units
of $a$, momenta in units of $[a (mk_L)^{1/2}]$, temperature in units
of $[k_L a^2/k_B]$, spring constants in units of $k_L$, frequencies
in units of $[(m/k_L)^{1/2}]$ and energies in units of $[k_L a^2]$.
In particular, we set in our simulations
$k_L=1,V_L=5,k_{int}=0.05,k_R=0.2$. For a typical situation, the
dimensionless temperature is set at $T_0 =0.09$. This yields with
$k_La^2/k_B \sim 10^3 K - 10^4 K$ a  physical temperature of the
order $T_0 \sim 90 K-900 K$.
%The choice of $T_0=0.09$ in simulation will guaranty that the real
%temperature is on the order of room temperature.
The equations of motion are integrated by the symplectic velocity
Verlet algorithm with a small time step $h=0.005$.  The system is
simulated for a total time $tt=2 \cdot 10^8$. The chosen optimal
coupling strength of the heat bath is fixed at $\eta=0.5$.

%Figure 2

\begin{figure}
\includegraphics[width=\columnwidth]{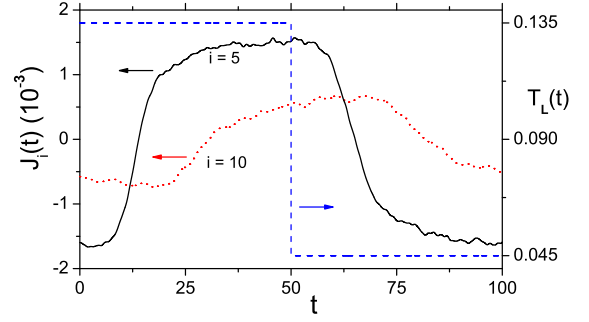}
\vspace{-.5cm} \caption{(Color Online) The numerically ($10^6$
realizations) evaluated, periodically varying asymptotic local heat
flux $J_i(t)$ over the full driving period $2\pi/\omega = 100 $ is
compared with the rocking bath temperature $T_L(t)$,  at driving
strength $A=0.5$
 and zero bias $\Delta=0$ for $T_0=0.09$. The lattice
 length is $N=50+50$.
 The numerically cycle averaged, spatially homogenous heat
 flux equals $J = -4.92 \cdot 10^{-6}$($i=5$) and $J= -4.93 \cdot 10^{-6}$ ($i=10$).  }
\label{fig:2}
\end{figure}

The  asymptotic   heat flux  $J_i(t)$ is assuming the periodicity of
the external driving period $2\pi/\omega$ after the transients have
died out. This fact is assured for all of our chosen frequencies
$\omega$ after a simulation time of $st \sim O(10^7)$. At those
asymptotic long times $t$ the heat flux equals  the noise average
$J_i(t) =k_{i} \left\langle\dot{q}_{i}(q_i-q_{i-1})\right\rangle$
where $k_i= k_L$ for $i=2,\dots,N/2$ and $k_i=k_R$ for
$i=(N/2)+2,\dots,N$, being here evaluated in the commonly employed
way, cf. in Refs.
\cite{rectifier,diode,Hu06,peyrard,yang,transistor,lepri}. The
static heat flux  $J$ then follows as the cycle average over a full
temporal period: $J=\frac{\omega}{2\pi} \int_{0}^{2\pi/\omega}
J_i(t) dt$, which with ergodicity being valid equals as well the
long time average, i.e. $J= k_{L}\overline{ \dot{q}_{2}(q_2-q_{1})}
= k_R \overline{\dot{q}_{N}(q_{N}-q_{N-1})}$. In Fig. \ref{fig:2} we
depict the resulting periodic  variation of the heat flux $J_i(t)$
{\it vs.} the externally applied temperature variation $T_L(t)$.
This nonlinear response  exhibits a characteristic phase lag {\it
vs.} the perturbation $T_L(t)$ and is dynamically  biased to yield a
nonzero temporal average.

%Figure 3
\begin{figure}
\includegraphics[width=\columnwidth]{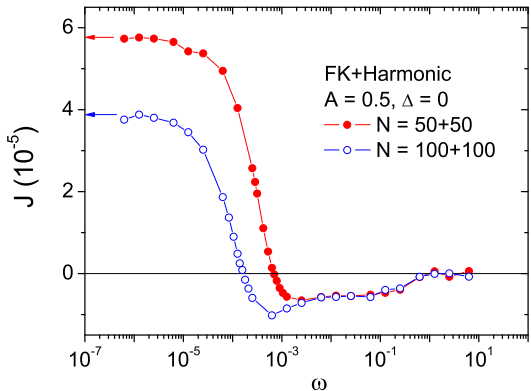}
\vspace{-.5cm} \caption{(Color-online) Time averaged heat flux $J$
{\it vs.} the angular driving frequency $\omega$ for different
lattice lengths $N=50+50$ and $N=100+100$ and $T_0=0.09$.  The two
arrows mark the heat flux  calculated in the adiabatic limit.}
\label{fig:3}
\end{figure}

With the introduction of a temporally alternating temperature field
in Eq. (\ref{T-setup}), we thus achieve a controllable manipulation
of heat flow by an externally adjustable parameter, i.e. the driving
frequency $\omega$. The static thermal bias $\Delta$ has been  set
to zero. The only resource  driving heat
 across the junction  thus is the non-equilibrium, alternating temperature field
$T_L(t)$, which generates positive and negative temperature
variations in the first and second half of driving period. As a
result, the direction of heat flow will tend to reverse each half
driving period. In the adiabatic limit; i.e. $\omega\rightarrow0$,
the alternating temperature $T_L(t)$ can be expressed by two
opposite static thermal bias values, yielding the average heat
current approximately as the averaged heat current for two opposite
static thermal bias values, see the two horizontal arrows depicted
in Fig. \ref{fig:3}.  In contrast, in the fast-driving limit
$\omega\rightarrow\infty$, the left-end atom will experience a
time-averaged constant temperature. This corresponds to thermal
equilibrium, yielding  $J\rightarrow0$ when
$\omega\rightarrow\infty$. In Fig. \ref{fig:3}, the numerically
determined average heat current is depicted  as a function of the
driving frequency $\omega$ for the driving amplitude $A=0.5$. In
full agreement with our predictions, a finite heat current $J$
emerges in the adiabatic limit $\omega\rightarrow0$, becomes
diminished in the non-adiabatic limit and essentially vanishes for
large $\omega$. At adiabatic driving the values of  $J$ agree well
with the numerical values determined from the adiabatic
approximation. A tantalizing observation is that $J$ does not vanish
monotonically. Remarkably, at some intermediate value $\omega$, the
heat flow $J$ crosses zero and subsequently  {\it reverses} its
direction upon further increasing $\omega$. Consequently, the
direction of net heat flow can be manipulated  by suitably tailoring
the frequency of the temperature variations.

This interesting reversal of the heat flux  can be related to the
thermal response time of the system. The non-rocked FK lattice obeys
Fourier's law \cite{HuLiZhao}. Thus, the corresponding temperature
variations $T(t,x)$ obey the diffusion equation:
$\partial{T}/\partial{t}=(\kappa/c)\partial^2{T}/\partial{x^2}$,
where $\kappa$ denotes the heat conductivity and $c$ is the specific
heat. The solution follows a Gaussian wave packet
$T(t,x)=1/2\sqrt{\pi\kappa t/c}\exp(-x^2c/4\kappa t)$. The thermal
response time can now be estimated as the time span for the energy
to diffuse across the system, i.e. $\tau\sim c N^2/4\kappa$. At
temperature $T_0= 0.09$, the FK lattice assumes the numerical values
 $\kappa\sim 0.5$ and $c\sim 1$. Thus, the characteristic frequency
scale $\omega_c$ of the system can be  estimated as
$\omega_c=2\pi/\tau$.  This characteristic frequency  scales
inversely with the system size $\propto N^{-2}$. For $N=100$ we then
find $\omega_c  \sim 10^{-3}$  and a roughly four times smaller
value for $N=200$. These two  estimates for $\omega_c$ are in good
agreement with those observed numerically  in Fig. \ref{fig:3}.
Taking the physical unit of frequency, i.e.  $\omega_0 =
[(m/k_L)^{1/2}] \sim 10^{13}\, \text{sec}^{-1}$  into account,
$\omega_c\sim 10^{-3}$ corresponds to a typical physical frequency
for microwaves of  $\sim 10^{10}\,\text{sec}^{-1}$. The
theoretically predicted
 red shift $\propto N^{-2}$ for $\omega_c$ with increasing
system size is nicely corroborated by our numerical results.

%Figure 4

\begin{figure}
\includegraphics[width=\columnwidth]{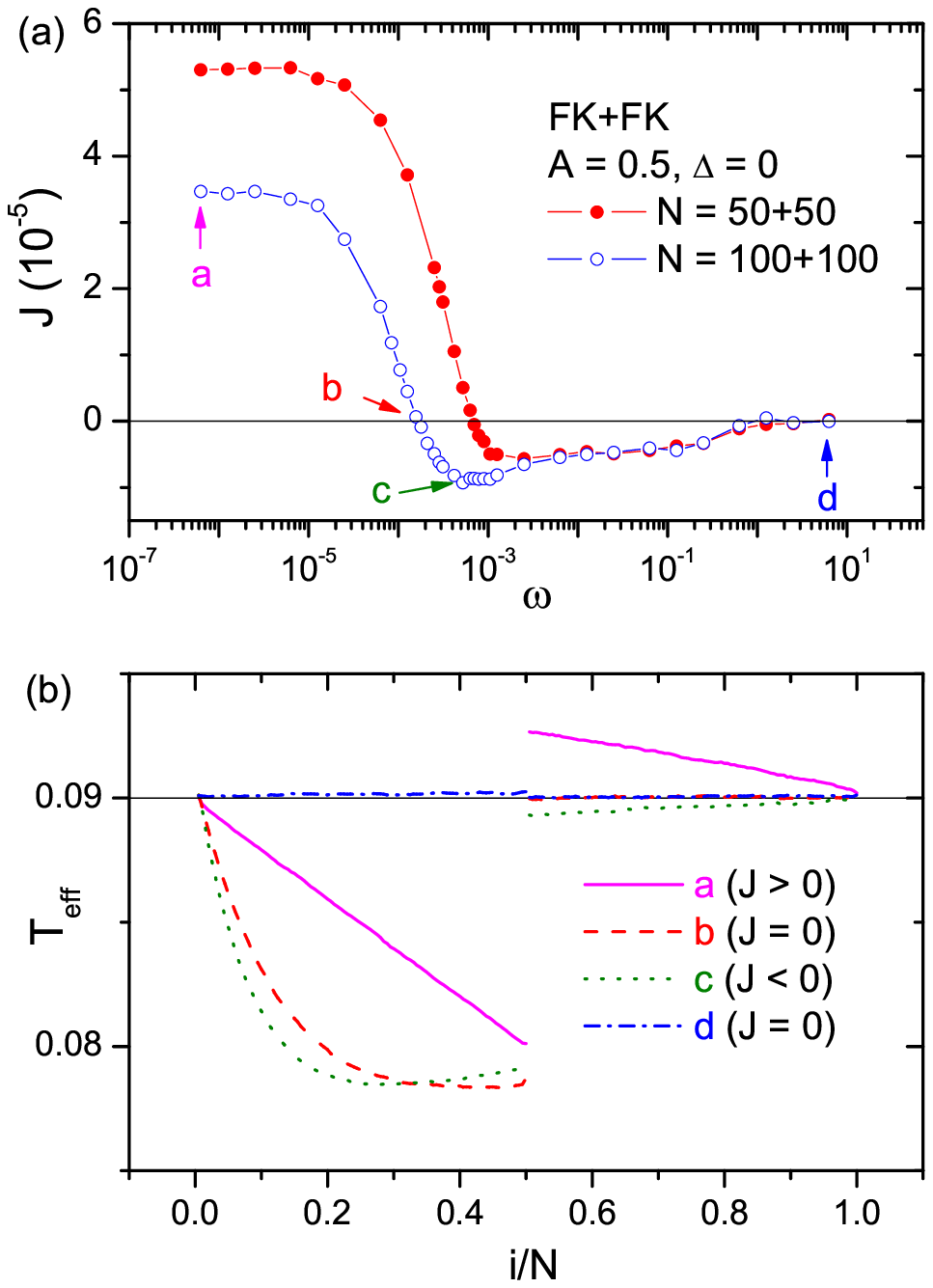}
\vspace{-.5cm} \caption{(Color-online) (a) Stationary heat flux $J$
{\it vs.} driving frequency $\omega$ for different lattice lengths
$N=50+50$ and $N=100+100$ at $T_0=0.09$ and a driving strength $A= 0.5$.
(b) The effective
temperature profiles of four selected points in (a) at length
$N=100+100$.} \label{fig:4}
\end{figure}

To gain additional insight into this reversal of heat flow   we
investigate  the local temperature variations across the junction at
different driving frequencies $\omega$. After evolving the system
over a long total simulation time $tt=2\cdot10^8$ this local
temperature of $i$-th atom is evaluated from its temporal long time
average of the kinetic energy, i.e. $T_{eff}(i)=
\overline{\dot{q}^2_i}$. In doing so, we switch to a FK-FK
configuration because the harmonic lattice knowingly is not able to
build up a temperature gradient \cite{lepri}. The employed right
sided FK lattice has the parameters $(k_R=0.2,V_R=1)$. In Fig.
\ref{fig:4}(a) a similar heat current modulation as for the FK-HL
configuration is observed for the FK-FK junction with $\Delta=0$ and
$A=0.5$. The effective temperature profiles of four numerical runs,
denoted as (a,b,c,d) in Fig. \ref{fig:4}(a),  are depicted in Fig.
\ref{fig:4}(b) versus the relative site positions $i/N$.

%Figure 5

\begin{figure}
\includegraphics[width=\columnwidth]{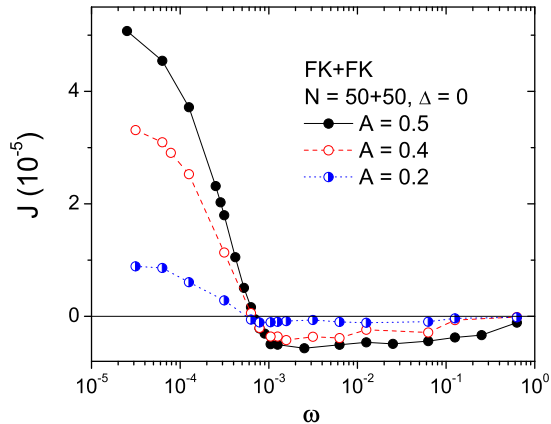}
\vspace{-.5cm} \caption{(Color-online) Stationary heat flux $J$ {\it
vs.} driving frequency $\omega$ for different driving strengths $A$
of the temperature modulation for a size $N=50+50$  at $T_0=0.09$
and zero thermal static bias $\Delta$.} \label{fig:6}
\end{figure}

In clear contrast to a non-rocking case  (i.e. $A=0$)   with no net
temperature bias, a distinct temperature gradient now emerges for  a
rocking temperature $T_L(t)$. The temperature profile exhibits a
discontinuity at the interface. For the situation in (c) where the
heat flow reversal $J$ approaches  its lowest, negative value, the
effective temperature profile becomes rather complicated: Away from
the bending part (with a negative-valued slope) towards the left
terminal side the resulting temperature profile exhibits an opposite
thermal gradient in comparison to case (a)--thus indicating that a
current reversal occurs. This opposite thermal gradient behavior can
also be detected upon comparing the temperature gradients of case
(b) and (d), exhibiting both a vanishing heat flow $J=0$. The
discontinuity of $T_{eff} (i)$ occurring at the interface reaches
its maximal value at low frequencies, cf. case (a),  and
increasingly diminishes with increasing driving frequency beyond the
frequency value for reversal, cf. case (d).

%%%% New %%%%%%%%%%%
We next investigated numerically the role of the driving strength
$A$ of the temperature modulation versus driving frequency $\omega$.
The results are depicted with Fig. (\ref{fig:6}). As expected, a
lower driving strength consistently yields smaller values of the
Brownian motor induced heat flux $J$, which vanishes identically
when the strength of the source of nonequilibrium is approaching
zero, i.e. $A=0$. Because the  frequency scale $\omega_c$  for
occurrence of heat flux reversal is mainly size dependent $\propto
N^{-2}$ we expect a weak dependence of $\omega_c$ on driving
strength $A$. This result is corroborated with Fig. (\ref{fig:6})
where this characteristic frequency is practically independent of
driving strength $A$.
%%%%%%%%%%%%%%%%%%%%%%%%%%%%%%%%

\begin{figure}
\includegraphics[width=\columnwidth]{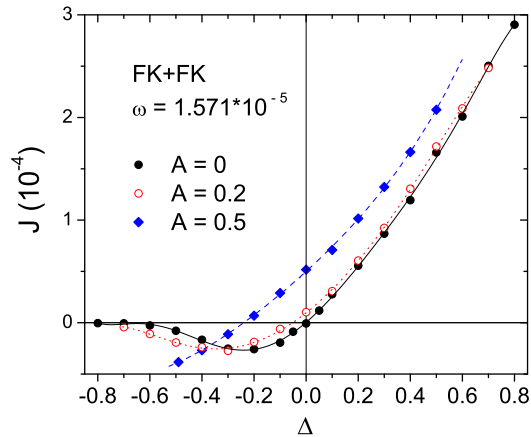}
\vspace{-.5cm} \caption{(Color-online) Heat flux $J$ {\it vs.}
thermal bias $\Delta$ for different driving amplitudes $A=0,\, 0.2$,
and $0.5$. The lattice length is $N=50+50$ and $T_0=0.09$. Note that
the nonlinearity in the Frenkel-Kontorva part of the junction is
essential to obtain the thermal ratchet effect. At large rocking
strength ($A=0.5$) the current-bias characteristics can be
manipulated to eliminate a NDTR-regime at negative bias values
$\Delta$. } \label{fig:5}
\end{figure}

{\it Flux-bias characteristics}. The  finite ratchet value of heat
flux $J$ in the absence of static thermal bias $\Delta$ now allows
for directing heat current {\it against} a non-zero thermal bias
$\Delta$. In Fig. \ref{fig:5}, we depict the flux-bias
characteristics $J-\Delta$ at small driving frequency
$\omega=1.571\cdot 10^{-5}$ ($\sim 100$ MHz), which is in the range
of ultrasonic frequencies. The $J-\Delta$ curve in absence of
time-dependent manipulation ($A=0$) is presented as a reference,
obeying $J(\Delta=0)=0$, in agreement with the $2$nd Law. The dashed
line corresponds to a driving amplitude $A=0.5$. We now detect a
finite heat current $J$ at {\it vanishing} thermal bias $\Delta=0$.
For negative  bias within the range $\Delta\in[-0.23,0]$, the
direction of the heat flow is positive. With $J$ taken as the
average over a driving period this implies that heat flows from cold
to hot. The so called ``stall bias" of the thermal bias that yields
$J=0$ is located around $\Delta=-0.23$. Moreover, we detect that the
characteristics of (NDTR) becomes modified as well by the switch-on
of the alternating bath temperature $T_L(t)$. The original working
range of NDTR at $A=0$ is $\Delta\in[-0.6,-0.2]$. At $A=0.2$, this
range undergoes a shift towards $\Delta\in[-0.7,-0.3]$. For $A=0.5$,
this very NDTR-phenomenon effect can be {\it dynamically eliminated}
all together. This effect is thus of prominent interest for the
design of a efficient thermal transistor: It allows one to change
the range of working temperatures of a thermal transistor by solely
adjusting the strength of the driving temperature field.

%Figure 5

In conclusion, we have shown that heat flow across a structured
nonlinear lattice junction can be efficiently  controlled by use of
a temporally alternating temperature field. The heat flow becomes
directed  and even can be reversed by suitably selecting the driving
frequency. In clear contrast to the thermal diode effect, the
presented Brownian heat rachet physics dynamically generates a
finite {\it heat flux} at zero thermal bias $\Delta=0$. Thus, we
deal with a new phenomenon which, as well, is in distinct contrast
to the by now common situation of thermally assisted, directed {\it
particle transport} in Brownian motors \cite{BM1,BM2,BM3,BM4}. This dynamically
induced heat flow can be directed against a non-zero, time-averaged
net thermal bias $\Delta$. This fact, however, does not necessarily
imply an active overall cooling of the device, cf. also Fig.
\ref{fig:4} (b). The size and the shape of NDTR can be manipulated
as well. Nonlinearity, as reflected with the FK-segment, is
essential for the phenomenon: a junction composed of two asymmetric
harmonic lattices fails to support a rachet heat flux. All these
phenomena call for beneficial applications in the control and
management of heat flow on the micro- and nano-scale. With a future,
more detailed work \cite{future} we  shall investigate, how the size
of the directed heat flux can be enlarged by stylizing further the
nonlinearity of the lattice structure. In particular, we expect much
larger heat currents when using a Fermi-Pasta-Ulam chain containing
both, a cubic and a quartic interaction potential.

Our setup  constitutes a two-segment rectifier model \cite{diode}
and the driving frequency is in a range of ($\lesssim 100$ MHz). The
finite, directional heat flux is feasibly detected by slowly rocking
the bath temperature within the adiabatic regime. Given the fact
that rectification has been demonstrated experimentally in
asymmetrically deposited nanotubes \cite{experimentaldiode}, we are
confident that the appealing heat ratchet phenomena presented herein
will invigorate experimentalists to validate our findings by
designing such thermal ratchet systems.

\acknowledgments The authors like to thank Prof. L. Wang for his
insightful comments on this work. The work is supported in part by
an ARF grant, R-144-000-203-112,
 from the Ministry of Education of the Republic of
Singapore, grant R-144-000-222-646 from NUS and by the German
Excellence Initiative via the \textit {Nanosystems Initiative
Munich} (NIM) (P.H.).


\begin{thebibliography}{01}
\bibitem{CStarr} C. Starr, J. Appl. Phys. \textbf{7}, 15 (1936).

\bibitem{Williams66} A. Williams, Ph.D Thesis, Manchester University
(1966).

\bibitem{Eastman68} G. Y. Eastman, Sci. Am. {\bf 218}, 38 (1968).

\bibitem{Thomas70} T. R. Thomas and S. D. Probert, Int. J. Heat Mass
Transfer, {\bf 12}, 789 (1970).

\bibitem{rectifier70} P. W. O'Callaghan, S. D. Probert, and A. Jones, J. Phys. D {\bf 3}, 1352 (1970).

\bibitem{rectifier} M. Terraneo, M. Peyrard, and G. Casati, Phys. Rev. Lett. \textbf{88}, 094302 (2002).

\bibitem{diode} B. Li, L. Wang, and G. Casati, Phys. Rev. Lett. \textbf{93}, 184301 (2004).

\bibitem{quantumdiode} D. Segal and A. Nitzan, Phys. Rev. Lett. \textbf{94}, 034301 (2005).


\bibitem{Hu06}B. Hu, L. Yang, and Y. Zhang, Phys. Rev. Lett. \textbf{97}, 124302
(2006).
%%%%%%%%%%new
\bibitem{peyrard}
M.~Peyrard, Europhys. Lett. \textbf{76}, 49 (2006).
%%%%%%%%%%%%%%%

\bibitem{yang}N. Yang, N. Li, L. Wang, and B. Li, Phys. Rev. B {\bf 76}, 020301(R) (2007).

\bibitem{experimentaldiode} C. W. Chang, D. Okawa, A. Majumdar, and A. Zettl, Science \textbf{314}, 1121 (2006).

\bibitem{transistor} B. Li, L. Wang, and G. Casati, Appl. Phys. Lett. \textbf{88}, 143501 (2006).


\bibitem{WangLi07} L. Wang and B. Li, Phys. Rev. Lett. {\bf 99},
177208 (2007).

\bibitem{WangLi08} L. Wang and B. Li, Physics World {\bf 21} (3),
28 (2008).


\bibitem{BM1}
P. Reimann {\it et al.}, Phys. Lett. A {\bf 215}, 26 (1996).
\bibitem{BM2}
R. D. Astumian  and P. H\"anggi, Phys. Today {\bf 55} (11), 33 (2002).
\bibitem{BM3}
P. H\"anggi, F. Marchesoni, and F. Nori, Ann. Phys. (Leipzig) {\bf 14},
51 (2005).
\bibitem{BM4}
P. Reimann and P. H\"anggi, Appl. Phys. A \textbf{75},
169 (2002).

%%%%% new
\bibitem{vandenbroeckPRL2006}
C. Van den Broeck, Phys. Rev. Lett. \textbf{96}, 210601 (2006).
\bibitem{segalPRE2006}
D. Sigal and A. Nitzan, Phys. Rev. E {\bf 73}, 026109 (2006).
\bibitem{marathePRE2007}
R. Marathe, A. M. Jayannavar and A. Dhar, Phys. Rev. E {\bf 75},
030103(R) (2007).
\bibitem{vandenbroeckPRL2008}
M. van den Broeck and C. Van den Broeck, Phys. Rev. Lett. {\bf 100}, 130601 (2008).

%%%%%%%%%%%%%%%%%%%%%%%%

\bibitem{lepri} S. Lepri, R. Livi, and A. Politi, Phys. Rep. {\bf
377}, 1 (2003).

\bibitem{HuLiZhao}
B. Hu, B. Li, and H. Zhao, Phys. Rev. E \textbf{57}, 2992 (1998).

\bibitem{Thompson} C.J.~Thompson, {\it Classical Equilibrium
Statistical Mechanics}, (Calendron Press, Oxford 1988).

\bibitem{future}
N. Li, P. H\"anggi, and B. Li, extended version, in preparation.


\end{thebibliography}
\end{document}